\documentclass{ws-p9-75x6-50}
\input psfig.tex
\begin{document}

\title{Relation of Light Curve Behaviour with Accretion Rates in Black Hole Candidate GRS1915+105}

\author{Sandip K. Chakrabarti, A. Nandi and S. G. Manickam}

\address{S.N. Bose National Centre for Basic Sciences, JD-Block, Salt Lake, Calcutta
700098\\E-mail: chakraba@boson.bose.res.in, anuj@boson.bose.res.in and sivman@boson.bose.res.in}

\maketitle

\abstracts{We report some properties of the light curves of GRS 1915+105 and 
present their interpretation using the variation of net accretion rate.}

\noindent  Proceedings of 9th Marcel grossman Conference (Ed. R. Ruffini)

Recently, Belloni et al\cite{belo00} classified all light curves based on energy channels available in RXTE data. 
However, one needs to use channels dictated by physical considerations. When matter accretes in Keplerian 
and sub-Keplerian components, the centrifugal barrier heats up and puffs up the disk close to the hole.
The hot electrons Comptonize intercepted soft photons ($\sim 2-3$keV) emitted by the Keplerian disk. 
Power-law hard photons pivot at $ \sim 17-18$ keV and the number of photons with
even higher energy is fewer. Thus, we separate the photons in A=$0-3$keV, B=$3-17$keV
and C=$17-60$keV. Fig. 1 shows all types of `softness' ratio diagrams plotted with 
$A/C$ along Y-axis and $B/C$ along the X-axis\cite{nan00}. The numbered sequence is same as the sequence\cite{belo00}
$\chi,\ \alpha, \ \nu, \ \beta, \ \lambda,\  \kappa, \rho, \ \mu, \ \theta, \ \delta, \gamma$ and $\phi$
respectively. Here, Y-axis roughly measures the accretion rate of the Keplerian
disk and X-axis roughly measures a combination of the net accretion rate and geometry of the puffed up region
(actually, soft photons intercepted by it). Our softness ratio is
basically a collection of straight lines with slopes varying from class to class, clearly indicating 
that two component flows are essential\cite{ct95} and 
that the Keplerian and sub-Keplerian flows are not of constant ratio. Spectrum for 
cases in Panel 1 is Hard, that of Panel 9 is Semi-Soft, those of Panels 10-12 are Soft
and the rest belongs to Intermediate classes where transitions from low to high count take place\cite{nan00}. 

\begin{figure}
\vbox{
\vskip -1.0cm
\hskip 0.0cm
\centerline{
\psfig{figure=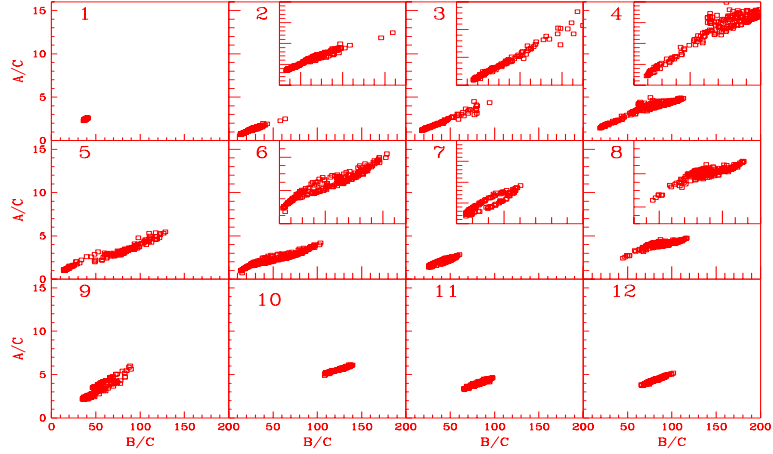,height=12truecm,width=11truecm}}}
\vspace{-4.7cm}
\caption[] {Softness Ratios A/C vs. B/C  plotted for all the 12 types of light curves.}
\end{figure}

\begin{figure}
\vbox{
\vskip -1.5cm
\hskip 0.0cm
\centerline{
\psfig{figure=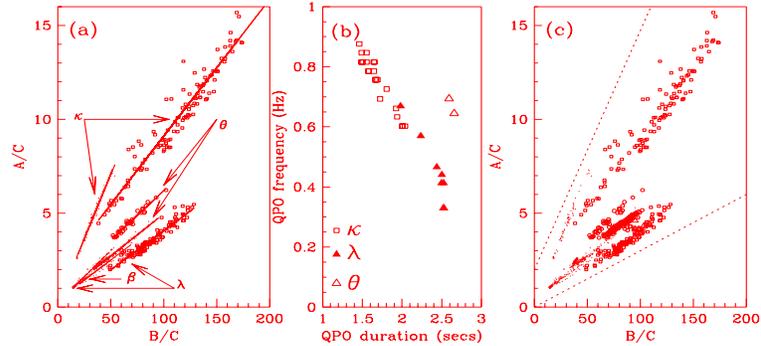,height=13.0truecm,width=11.5truecm}}}
\vspace{-7.5cm}
\caption[] {(a) Softness ratio diagrams using our `paradigm' recommended channels for
$\kappa$, $\beta$, $\lambda$ and $\theta$ classes of Belloni et al\cite{belo00}. (b) Chakrabarti-Manickam
correlation for these classes show progressive rise of normalization from $\theta$ to $\lambda$ to $\kappa$
types. (c) State-wise separation of the parameter space.}
\end{figure}

Belloni et al\cite{belo00} pointed out that there are three basic states (A, B and C).
Following Chakrabarti et al\cite{chaknan,cetal00} we rename them  to States HW (Hard state with Winds), 
CC (Compton cooled sonic sphere) and EA (Enhanced Accretion) respectively.
In each class of the light curve, several States
may be observed. While these states correspond to three `hazy' areas
(banana, atoll, Z-shaped etc.)
in hardness ratio diagram\cite{belo00}, according to our choice, 
for a given class, they roughly correspond to straight
lines (Fig. 1), only slopes vary from class to class.
In Fig. 2a, we fit lines in classes $\kappa$, $\lambda$, $\beta$
and $\theta$. Dots represent HW (C\cite{belo00})
and circles represent CC (A) and squares are mixtures of CC (A) and EA (B). Slopes of straight lines in
HW states in $\kappa$, $\lambda$ and $\theta$ are $0.137$, $0.055$ and $0.045$ respectively.
In Fig. 2b, correlation of duration of QPOs 
with QPO frequencies are plotted\cite{cm00} for $\theta$, $\kappa$ and $\lambda$ classes.
When fitted\cite{cm00} the accretion factor $\Theta_{dot m}$ follows the same sequence as the slopes given
above. Thus these two figures {\it independently} give handles on accretion rates.
In Fig. 2c, dots, small circles, triangles and squares represent HW (C), CC (A),
EA (B) and mixtures of CC (A) and EA (B) respectively. Dotted boundaries confine
softness ratios. Variation of slopes of various classes of light curves
in a given state is the proof that a pure Keplerian disk, or a pure
ion pressure supported Compton cloud cannot explain the observations and one definitely requires
two component accretions flows\cite{ct95} as demanded by advective disk paradigm.

\end{document}